\documentclass[prb,twocolumn,showpacs,showkeys,preprintnumbers,superscriptaddress,amsmath,amssymb]{revtex4-1}
\usepackage{graphicx}
\usepackage{epstopdf}
\usepackage{longtable}
\usepackage{dcolumn}
\usepackage{bm}
\usepackage{amsmath}
\usepackage{xcolor}
\usepackage{times}

\newcommand{\moseries}{Cu$_{1-x}$Mo$_{x}$Sr$_2$(Y,Ce)$_s$Cu$_2$O$_{5+2s+\delta}$}
\newcommand{\mofirst}{Cu$_{1-x}$Mo$_{x}$Sr$_2$YCu$_2$O$_{7+\delta}$}
\newcommand{\mohpo}{Cu$_{0.75}$Mo$_{0.25}$Sr$_2$YCu$_2$O$_{7.54}$}

\begin{document}

\title{Bulk superconductivity at 84 K in the strongly overdoped regime of cuprates}

\author{A. Gauzzi}
\email[Corresponding Author.~E-mail:]{andrea.gauzzi@upmc.fr}
\author{Y. Klein}
\affiliation{IMPMC, Sorbonne Universit\'es-UPMC, CNRS, IRD, MNHN, 4, place Jussieu, 75005 Paris, France}
\author{M. Nisula}
\author{M. Karppinen}
\affiliation{Aalto University, Department of Chemistry, Espoo 00076, Finland}
\author{P. K. Biswas}
\altaffiliation[Present address: ]{ISIS Facility, Rutherford Appleton Laboratory-STFC, Chilton, Didcot, OX11 0QX, United Kingdom.}
\author{H. Saadaoui}
\author{E. Morenzoni}
\affiliation{Laboratory for Muon Spin Spectroscopy, Paul Scherrer Institut, CH-5232 Villigen PSI, Switzerland}
\author{P. Manuel}
\author{D. Khalyavin}
\affiliation{ISIS Facility, Rutherford Appleton Laboratory-STFC, Chilton, Didcot, OX11 0QX, United Kingdom}
\author{M. Marezio}
\affiliation{Institut Ne\'el, CNRS and Universit\'e Grenoble Alpes, BP 166, 38042 Grenoble, France}
\author{T. H. Geballe}
\affiliation{Stanford University, Department of Applied Physics \& Materials Science, Stanford CA 94305, USA}

\date{\today}

\begin{abstract}
By means of magnetization, specific heat and muon-spin relaxation measurements, we investigate high-pressure oxidized \mohpo, in which overdoping is achieved up to $p \sim 0.45$ hole/Cu, well beyond the $T_c - p$ superconducting dome of cuprates, where Fermi liquid behavior is expected. Surprisingly, we find bulk superconductivity with $T_c$=84 K and superfluid density similar to those of optimally doped Y123. On the other hand, specific heat data display a large electronic contribution at low temperature, comparable to that of nonsuperconducting overdoped La214. These results point at an unusual high-$T_c$ phase with a large fraction of unpaired holes. Further experiments may assess the Fermi liquid properties of the present phase, which would put into question the paradigm that the high $T_c$ of cuprates originates from a non-Fermi liquid ground state.  
\end{abstract}




\maketitle
It is widely accepted that the familiar dome-like dependence of the superconducting critical temperature, $T_c$, upon doped hole concentration, $p$, in the CuO$_2$ plane observed in most cuprates \cite{zha93,tal95} reflects a connection between superconductivity and antiferromagnetism. $T_c$ vanishes for $p \lesssim 0.08$ hole/Cu, as the density of mobile holes and hence the superfluid density \cite{eme95} approaches zero. $T_c$ vanishes again upon overdoping for $p$'s larger than a critical value, $p_c \approx 0.27$ hole/Cu, where the strength of the antiferromagnetic correlations also vanishes \cite{bir88}. A further notion, which is generally assumed to be valid is that the overdoped materials with $p > p_c$ behave more or less as conventional metals described by Fermi liquid theory. In fact, this second assumption is not supported by much empirical data and it may not be correct for the first ($s = 1$) member of the homologous seeries of cuprates \moseries, originally synthesized by Ono \cite{ono93} and then synthesized and characterized together with the higher members of the series up to $s = 6$ by other groups \cite{mor04,gri06,gri10,chm10,marezio13,marezio14}, \footnote{Non-single-phase compounds of the homologous series were also reported by Marik \textit{et al.} \cite{mar13,mar15} perhaps because Mo metal rather than MoO$_3$ has been being used as precursor, which could explain why the samples were found to contain Mo$^{5+}$.}. The high $T_c$'s up to 87 K observed in the $s = 1$ member in the strongly overdoped region well beyond the dome put into question the paradigm that the high $T_c$ of cuprates originates from a non-Fermi liquid ground state \cite{and96}.

In order to address this issue, in the present paper we report a systematic investigation by means of neutron diffraction, magnetization, susceptibility, heat capacity and muon-spin relaxation spectroscopy ($\mu$SR) on newly synthesized powders of the first ($s$=1) member, \mofirst,with $x$=0.25. The structure of this phase, commonly known as (Cu,Mo)1212, is closely related to the well known chain coumpound yttrium barium copper oxide (Y123) with the following differences (see \cite{chm10}, the inset of Fig. 3 and the Supplementary Information): 1) the Ba cation is replaced by the isovalent and smaller Sr ion; 2) a fraction $x$ of the Cu ions in the chain layers is replaced by Mo; 3) the oxygen atoms in the chain layers are randomly distributed along the $a$ and $b$ directions. Consequently the structure is tetragonal $P4/mmm$ rather than orthorhombic $Pmmm$ as in optimally doped Y123. The higher members are obtained by inserting $s-1$ blocks of (Y,Ce)O$_2$ between the Y and CuO$_2$ layers into the (Cu,Mo)1212 structure.

The (Cu,Mo)1212 powder samples object of this work were newly synthesized using the citrate sol-gel method and a modified oxidation treatment. Appropriate amounts of Y$_2$O$_3$, CuO, SrCO$_3$ and MoO$_3$ were dissolved in 1M HNO$_3$ solution. Citric acid and ethylene glycol were added and the solution was heated in oven at 200 $^{\circ}$C until a gel was formed. The obtained gel was then dried, powdered and calcined at 600 $^{\circ}$C for 12 h. The resulting powder was pressed into a pellet and annealed at 980 $^{\circ}$C for 2 $\times$ 24 h with intermediate grinding to obtain the as synthesized product. The oxidation was carried out by mixing the as synthesized powder with 35 mol \% of KClO$_3$ and the mixture was then subjected to high pressure treatment in a cubic-anvil type high-pressure apparatus at 4 GPa and 500 $^{\circ}$C. The powders were first studied by means of neutron diffraction at the WISH instrument of the ISIS facility at the Rutherford Appleton Laboratory. Rietveld refinement of the neutron diffraction data yields an oxygen content $\delta = 0.54$. The only impurity phases detected in the sample are the electronically inactive KCl (4.5 \%) and KClO$_3$ (0.5 \%) arising from the oxydizing agent. The very low background noise of the WISH diffractometer \cite{cha11} used for the present study enables us to rule out impurity concentrations $> 1 $\%.

\begin{figure}[h]
\centering
\includegraphics[width=\columnwidth]{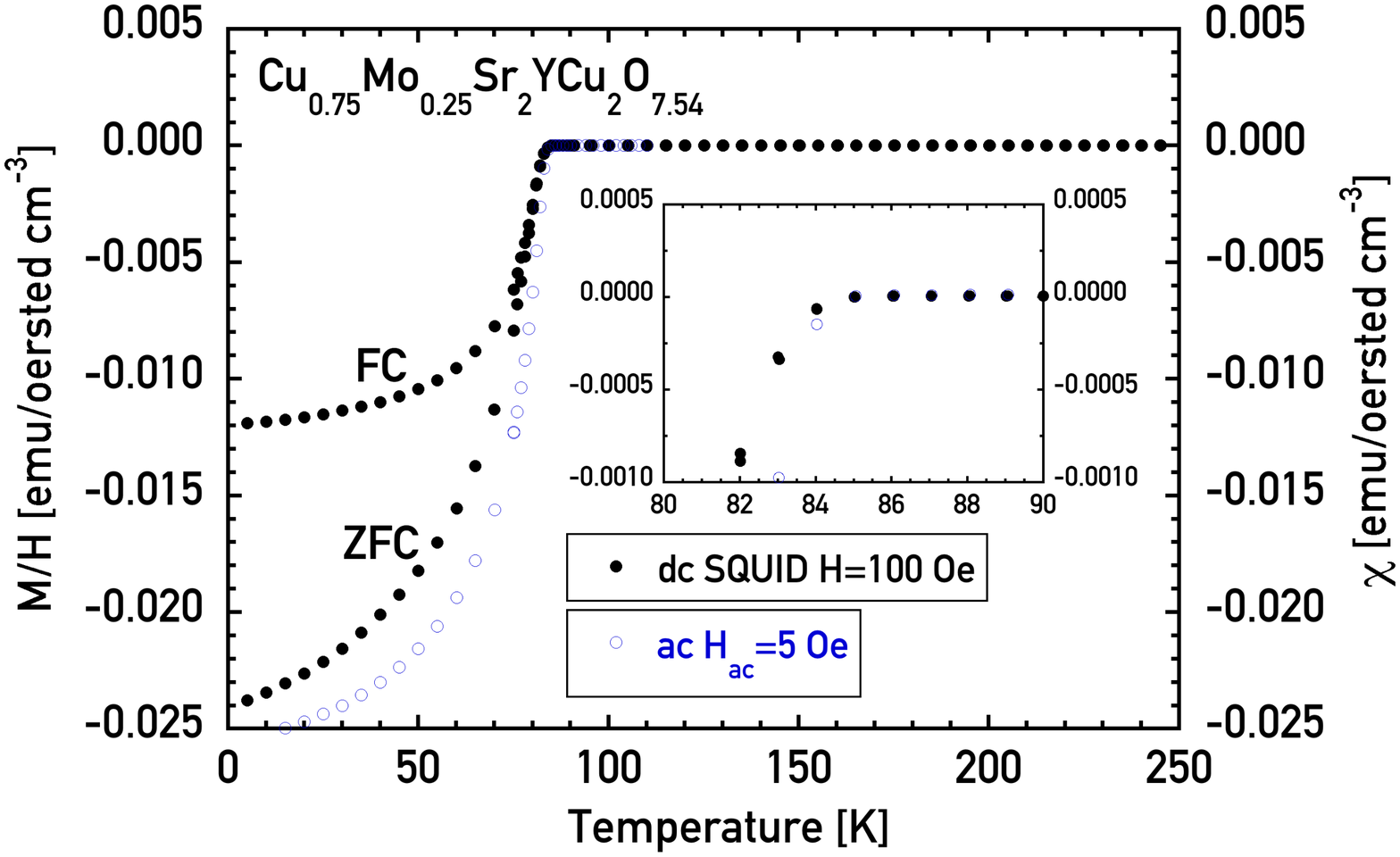}
\caption{\label{fig:susc} Zero-field- and field-cooling (ZFC, FC) magnetization (full symbols) and ac susceptibility $\chi$ (open symbols) of high-pressure oxidized (Cu,Mo)1212 powders. The two curves are compared using the same scale as the small dc field of 100 Oe allows to approximate $M/H$ with $\chi$. The inset shows a close-up of the transition.}
\end{figure}

\begin{figure}[h]
\centering
\includegraphics[width=\columnwidth]{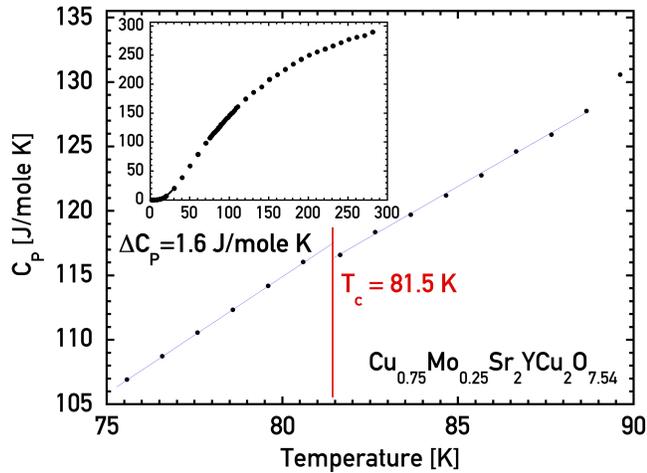}
\caption{\label{fig:hp} Isobaric specific heat, $C_p(T)$, measured on the same (Cu,Mo)1212 powder sample of Fig. 1. The main panel is a close-up of the transition showing the jump at $T_c$. The inset shows the behavior in the whole 2-300 K temperature range measured.}
\end{figure}

The superconducting transition of the powders was studied by means of dc magnetization measurements as a function of temperature in both zero-field (ZF) and field-cooling (FC) modes at 100 Oe using a commercial SQUID Quantum Design apparatus. In order to estimate more precisely the superconducting fraction, complementary measurements of ac magnetic susceptibility at a weak $H$=5 Oe field were also carried out using a commercial PPMS Quantum Design apparatus. A complementary study of the superconducting transition and of the electronic spectrum was carried out by means of specific heat measurements in the 2-300 K range performed using the relaxation rate method implemented in the same PPMS apparatus. Finally, the local properties of the superconducting state were probed by means of muon-spin relaxation spectroscopy. The experiments were performed on the GPS and Dolly instruments at the Swiss Muon Source S$\mu$S of the Paul Scherrer Institute. $\mu$SR is a powerful technique to determine the effective superfluid density and its temperature dependence. By applying a magnetic field perpendicular to the initial muon spin polarization, the inhomogeneous field distribution in the vortex state of the superconductor produces a damping of the $\mu$SR precession signal expressed by the muon spin relaxation rate $\sigma=\gamma_{\mu} \sqrt{\langle\Delta B^2\rangle}$, where $\gamma_{\mu}$ is the muon gyromagnetic ratio. From the width of the magnetic field distribution in the vortex state, $\sqrt{\langle\Delta B^2\rangle}$, the magnetic penetration depth, $\lambda$, and its temperature dependence are determined, which is related to the effective superfluid density, $\rho_s=\frac{n_s}{m^*}\propto \lambda^{-2} \propto \sigma$, where $n_s$ is the density of superconducting electrons and $m^{\ast}$ is their effective mass. Muon spin rotation spectra in the vortex state were taken as a function of increasing temperature by initially field-cooling the sample down to 1.6 K in different fields and analyzed assuming a Gaussian field distribution with relaxation rate $\sigma$.

\begin{figure}[h]
\centering
\includegraphics[width=\columnwidth]{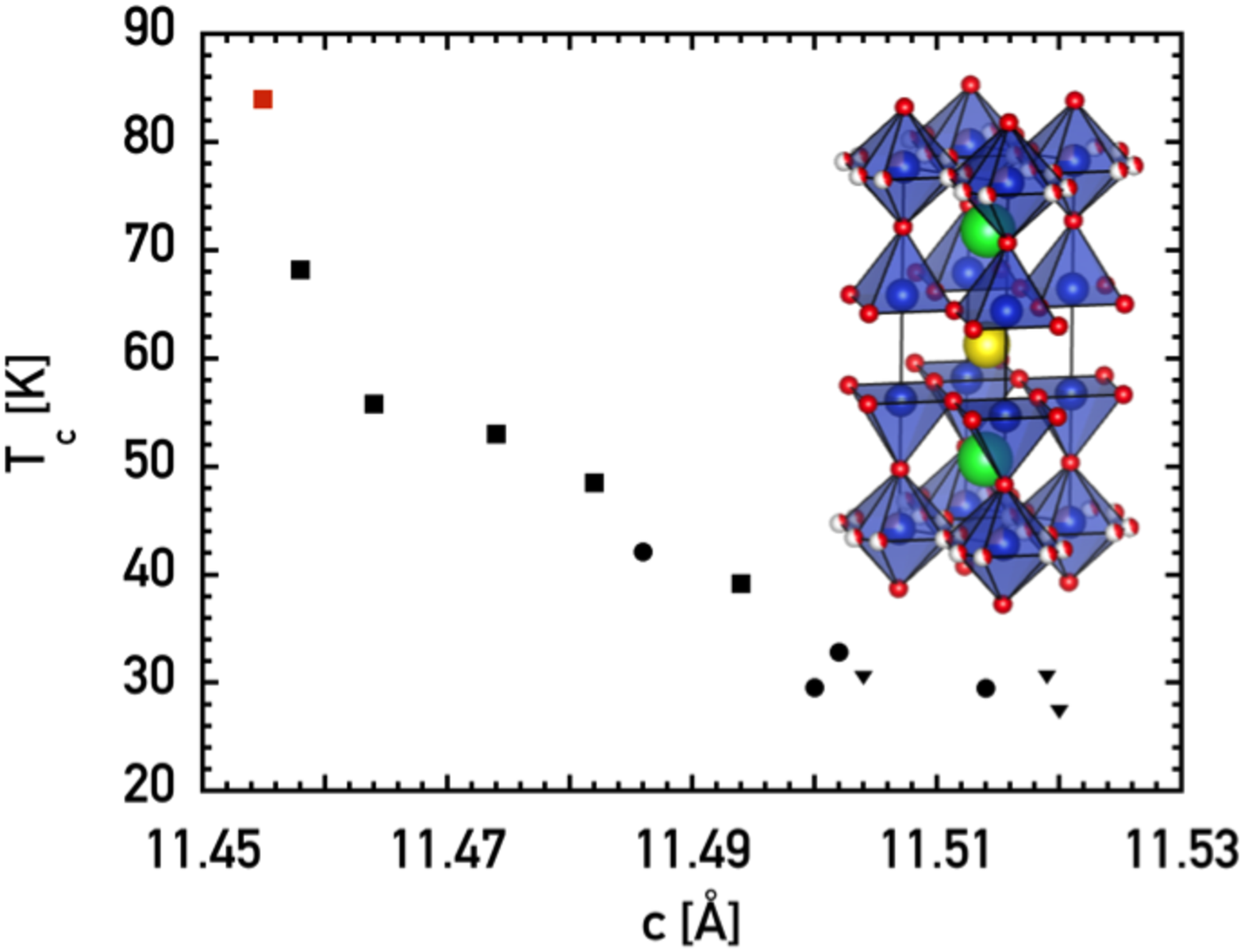}
\caption{\label{fig:Tcvsc} Correlation between $T_c$ and the $c$-axis parameter for a series of \mofirst~ samples prepared under different conditions. The red filled square corresponds to the present result obtained on a $x = 0.25$ high-pressure (HP) oxidized sample. Black simbols indicate previous results by Ono \cite{ono93} on $x = 0.20$ annealed HP samples (squares), HP samples (circles) and starting samples prepared at ambient pressure (triangles). The crystal structure is shown in the inset; blue, green, yellow and red spheres indicate Cu/Mo, Sr, Y and oxygen atoms, respectively; white-red spheres indicate partially occupied oxygen sites in the basal plane. The structure is described in detail in \cite{chm10} and in the Supplementary Information section.}
\end{figure}

In Fig. 1, we report the results of the dc magnetization and ac susceptibility measurements. Note the good agreement between the ZFC curve of the dc and ac data. Moreover, the ZFC dc magnetization curve is in fact a FC curve due to a small residual field $\approx$2 Oe trapped in the superconducting magnet. Both data set yield a $\approx$30\% screening volume; the slightly smaller diamagnetic signal of the dc data is attributed to the comparatively higher field of 100 Oe used, which leads to a larger flux penetration. As discussed in the past \cite{kit91, all93}, the above shielding fraction is typical for single-phase superconducting cuprates that depends upon various effects including flux pinning, among others. In the present case, considering the absence of impurities, this fraction should be then regarded as a lower limit for the Meissner fraction. We conclude that superconductivity is a bulk phenomenon associated with the oxidized \mofirst~ phase. Further evidence of bulk superconductivity is given in Fig. 2, where one notes the characteristic jump, $\Delta C_p$, at $T_c$ of the specific heat measured on the same powder sample wih $T_c$=84 K of Fig. 1 and by the $\mu$SR results to be discussed below. From Fig. 2, one obtains $\Delta C_p/T_c \approx 20$ mJ/mol K$^2$, a value comparable with those previously reported on other cuprate superconductors \cite{har92}. For example, $\Delta C_p/T_c \approx$ 11 (50) mJ/mol K$^2$ in underdoped (optimally doped) Y123, with $T_c$ =60 (90) K. The slightly lower $T_c$ value of the specific heat results is attributed to the lower accuracy within $\pm 1$ K inherent to the relaxation method used. 

The present data enable us to extend to higher $T_c$'s the inverse dependence of $T_c$ as a function of $c$-axis parameter previously reported by Ono \cite{ono93} for a series of (Cu,Mo)1212 samples with a similar, $x = 0.20$, Cu/Mo substitution level and with different oxygen doping. Whilst this dependence is characteristic of cuprates, as it reflects the decrease of the copper-apical oxygen distance, $d_{ap}$, upon hole doping, in the present oxidized (Cu,Mo)1212 sample, this distance is strikingly short. The neutron diffraction data presented in the Supplementary Information section show that superconductivity is introduced by the above high-pressure oxygen treatment while $d_{ap}$ is reduced from 2.29 \AA~ to 2.165 \AA. For comparison, in Y123, $d_{ap}$ decreases from 2.47 \AA~ in the reduced insulating phase YBa$_2$Cu$_3$O$_6$ to 2.29 \AA~ in optimally doped YBa$_2$Cu$_3$O$_{6.95}$ \cite{cav90}. Such a record short 2.165 \AA~ distance concomitant to a very high $T_c$ is surprising in the following two aspects:

1) It indicates an exceptionally large hole concentration in the CuO$_2$ planes, well above the maximum found in Y123 and well inside the overdoped region of cuprates. Using a self-consistent bond valence sum (BVS) analysis, where a weighted value of the radius parameters, $R_0^{2+}=1.655$ \AA~ and $R_0^{3+}=1.735$ \AA~ for the Cu$^{2+}$ and Cu$^{3+}$ ions, respectively, has been used \cite{chm10}, we estimate a planar Cu valence $v_{\rm Cu}=2.40$. This estimate is fully consistent with the value $v_{\rm Cu}=2.46$ reported in previous structural \cite{chm10} and XANES \cite{gri10} studies on similar superconducting Mo-cuprate samples.

2) The occurrence of superconductivity with $T_c$ as high as 84 K concomitant to the above record short $d_{ap}$ is at odds with the empirical observation that cuprate families with higher $T_c$'s rather display longer $d_{ap}$'s \cite{oht91}, which has been supported by first principles calculations within the Hubbard model \cite{pav01,web12}. The present finding may therefore point at a new physics governing the strongly overdoped region of cuprates.

A picture of overdoping is supported by a straightforward analysis of the low-temperature behavior of the specific heat shown in Fig. 4. It is recalled that, in normal metals, the Sommerfeld constant, $\gamma$, experimentally measured as the residual value at $T$=0 K of the $C(T)/T$ vs $T^2$ curve is proportional to the carrier density, $n$. In a superconducting metal, $\gamma$, is expected to vanish at zero temperature and an estimate of $n$ from the specific heat requires a knowledge of the gap spectrum. In BCS $d$-wave superconductors like cuprates, where the existence of lines of nodes in the gap function, $\Delta_{\textbf k}$, give rise to a parabolic increase of the specific heat at low temperature \cite{mom96}. In this case, $\gamma$ is related to the curvature of the parabola and its determination requires a precise knowledge of the angular dependence of $\Delta_{\textbf k}$ and of the quasiparticle spectrum \cite{mom96,wan07}.

In the analysis of the present specific heat data, we adopt a simpler approach, for the low temperature $C(T)$ curve does not exhibit a clear parabolic dependence. The $C(T)/T$ vs. $T^2$ plot in Fig. 4 is flat at low temperature as if the gapless behavior was less pronounced than in the $d$-wave case. We then limit ourselves to compare the present data with those previously reported on optimally doped Y123 ($T_c$ = 92 K) \cite{mol97} and optimally- and over-doped La$_{2-x}$Sr$_x$CuO$_4$ (La214) single crystals \cite{wan07} with $T_c$ values ranging from the maximum $T_c \approx 35$ K to 0 K. Our data exhibit a very large residual electronic specific heat normalized to the number of planar Cu atoms, $\gamma \approx 10$ mJ K$^{-2}$ Cu mol$^{-1}$, about 4-5 times larger than the values reported on the above optimally doped systems. The present value is rather comparable to the values found in overdoped La214, with vanishing $T_c$, or even beyond in the metallic nonsuperconducting phase, where $\gamma$ probes the density of states of \textit{normal} electrons. This conclusion is supported by similar results on optimally doped Bi$_2$Sr$_2$CaCu$_2$O$_8$ \cite{jun94} and overdoped Tl$_2$Ba$_2$CuO$_{6+\delta}$ \cite{lor94}.

\begin{figure}[h]
\centering
\includegraphics[width=\columnwidth]{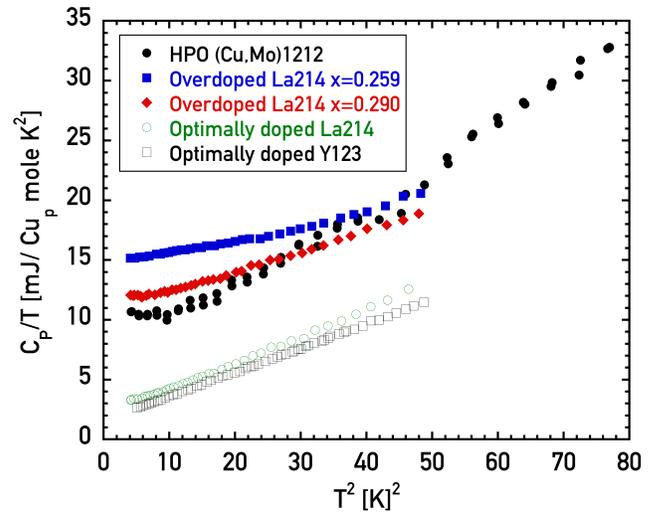}
\caption{\label{fig:gamma} Comparison of the low-temperature specific heat of the high-pressure oxydized (HPO) (Cu,Mo)1212 sample of the previous figure (black dots) with previously reported data on overdoped La$_{2-x}$Sr$_x$CuO$_4$ with $x$=0.259 ($T_c$=6.5 K, blue squares) and $x$=0.290 (non superconducting, red diamonds) and on optimally doped La$_{2-x}$Sr$_x$CuO$_4$ with $x$=0.178 (green open circles) and YBa$_2$Cu$_3$O$_{7-\delta}$ with $\delta$=0.05 (black open squares). La214 and Y123 data are taken from \cite{wan07} and \cite{mol97}, respectively.}
\end{figure}

\begin{figure}[h]
\centering
\includegraphics[width=\columnwidth]{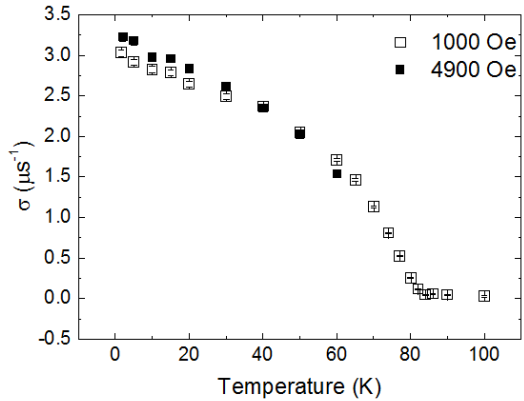}
\caption{\label{fig:muSR} Temperature dependence of the $\mu$SR rate, $\sigma$, at two different magnetic fields for an oxydized (Cu,Mo)1212 powder sample similar to that of Figs. 1-2.}
\end{figure}

As to the origin of such a large $\gamma$, we rule out extrinsic effects like contribution of metallic impurities, as their amount is below the detection limit $\sim$1 \% of powder neutron diffraction. Further evidence of an electronically homogeneous phase is given by the $\mu$SR data. The data were obtained on a (Cu,Mo)1212 powder sample prepared and oxydized under the same conditions as the previously discussed sample and with the same $T_c$=84 K. $\mu$SR is a phase volume sensitive technique and muons uniformly and locally probe the sample. So, if the sample contained different spatially separated phases, the muon signal would be the sum of different components, each component with its own precession frequency and broadening. In our sample we do not observe this behavior; the data can be fitted by a single precession signal with an amplitude corresponding to a 80\% superconducting fraction in the vortex state, a Gaussian relaxation rate $\sigma$ and a background contribution. Fig. 5 shows the temperature dependence of $\sigma$, which is proportional to the superfluid density. Note that measurements taken at different fields display a weak temperature dependence, as expected in an extreme type-II superconductor. At low temperature, $\sigma(0) = 3.0 \pm 0.1  \mu s^{-1}$, very close to that measured in optimally doped Y123 \cite{uem89}. At low temperatures, the linear decrease of $\sigma$ with temperature is also typical of a $d$-wave pairing symmetry. Overall the $\mu$SR data show that the sample is superconducting and homogeneous at least on length scales of $\sim 1 \mu$m corresponding to a few magnetic penetration depths. Thus, we cannot exclude inhomogeneities on much shorter length scales.

Having excluded the possibility of secondary phases or of other inhomogeneities on the microscopic length scale, the existence of a high temperature superconducting phase where simple metal physics is expected is our most important finding. In characterizing the superconducting state, several other striking features become apparent:

1) As in other clean cuprates, at low temperatures compared to $T_c$, the superfluid density decreases linearly with temperature. Generally, this is associated with the existence of nodal quasiparticles, as in the case of a clean, $d$-wave superconductor, which gives indirect evidence that even in this highly overdoped regime, the system is a $d$-wave superconductor.

2) Reconciliation of the $\mu$SR and heat capacity data puts tight constraints on an acceptable model. At the lowest temperatures we can probe, there remains a $T$-linear term in the specific heat with a finite intercept showing the presence of non-condensed holes in the superconducting state. In order to avoid proximity-induced condensation, this requires that these holes should be confined in regions larger than the coherence length, $\xi \sim$ 10 nm, but smaller than the penetration length, as discussed above. A scenario of a partial hole condensation in hole overdoped cuprates is supported by earlier specific heat data on overdoped La214 single crystals \cite{wan07} and, more recently, by kinetic inductance measurements on well characterized La214 single-crystal films, with $p = 0.295$, \textit{i.e.} at the edge of the dome \cite{boz16}.

The question arises whether the failure of complete condensation of the carriers in hole overdoped cuprates may be a more general phenomenon than has been realized. In order to verify this possibility, more studies on the (Cu,Mo)1212 phase are required. For example, it would be highly desirable to vary oxygen doping and to obtain single crystals, which are needed for transport and spectrocopic investigations. In alternative, we suggest that the epitaxial growth of thin films should be investigated.   

\begin{acknowledgments}
AG and MN acknowledge A. Daoud Aladine for his help in the structural refinement work. AG acknowledges fruitful discussions with M. Grilli, S. Caprara and Y.J. Uemura. THG is thankful for valuable inputs from S. A. Kivelson, S. Raghu and I. Bozovic during the course of this work, and for support by the DOE, Office of Basic Energy Sciences, under Contract No. DE-AC02-76SF0051.
\end{acknowledgments}

\bibliography{mo-cuprate}

\end{document}